\documentclass[twocolumn]{aastex63}
\usepackage{graphicx}
\usepackage{bm}
\usepackage{amssymb}
\usepackage{amsmath}
\usepackage{units}
\usepackage{array}\usepackage{lineno}
\usepackage{ulem}

\submitjournal{ApJ}

\usepackage{xcolor, lipsum}
\begin{document}

\title{Revisiting the Parameter Space of Binary Neutron Star Merger Event GW170817}

\correspondingauthor{Austin McDowell}
\email{atm426@nyu.edu}

\author{Austin McDowell}
\author{Andrew MacFadyen}
\affil{Center for Cosmology and Particle Physics, Physics Department, New York University, New York, NY 10003, USA}
\begin{abstract}
    Since the gravitational wave event GW170817 and gamma-ray burst GW170817A there have been numerous studies constraining the burst properties through analysis of the afterglow light curves. Most agree that the burst was viewed off-axis with a ratio of the observer angle to the jet angle ($\theta_{obs}/\theta_j$) between 4 - 6. We use a parameterized model and broadband synchrotron data up to $\sim 800$ days post-merger to constrain parameters of the burst. To reproduce the hydrodynamics of a gamma-ray burst outflow we use a two-parameter "boosted fireball" model. The structure of a boosted fireball is determined by the specific internal energy, $\eta_0$, and the bulk Lorentz factor, $\gamma_B(\sim 1/\theta_j)$ with shapes varying smoothly from a quasi-spherical outflow for low values of $\gamma_B$ to a highly collimated jet for high values. We run simulations with $\gamma_B$ in the range $1-20$ and $\eta_0$ in the range $2-15$. To calculate light curves we use a synchrotron radiation model characterized by $F_{peak}$, $\nu_m$, and $\nu_c$ and calculate millions of spectra at different times and $\theta_{obs}$ values using the \texttt{boxfit} radiation code. We can tabulate the spectral parameter values from our spectra and rapidly generate arbitrary light curves for comparison to data in MCMC analysis. We find that our model prefers a gamma-ray burst with jet energy $E_j\sim10^{50}$ ergs and with an observer angle of $\theta_{obs}=0.65^{+0.13}_{-0.14}$ radians and ratio to jet opening angle of ($\theta_{obs}/\theta_j$) = 5.4$^{+0.53}_{-0.38}$. 
\end{abstract}

\section{Introduction}

On August 17 2017 the LIGO-VIRGO consortium detected a gravitational wave (GW) signal, GW170817, from the first observed binary neutron star (BNS) merger (\cite{PhysRevLett.119.161101}). About 1.7 seconds later, the \textit{Fermi} Gamma-ray Burst Monitor detected a short gamma-ray burst (GRB), GRB170817A, consistent with the GW position
and after $\sim$11 hours the first optical signal was detected (\cite{2017ApJ...848L..14G}; \cite{2017Sci...358.1556C}). Follow-up observations in the x-ray, optical, and radio bands up to $\sim$1000 days after the merger revealed non-thermal synchrotron emission from the GRB afterglow (\cite{2017ApJ...848L..21A}; \cite{2017ApJ...848L..25H}; \cite{2017Sci...358.1579H}; \cite{2017ApJ...848L..20M}; \cite{2017Natur.551...71T}; \cite{2018ApJ...853L...4R}; \cite{2018ApJ...856L..18M}; \cite{2018ApJ...863L..18A}; \cite{2018ApJ...858L..15D}; \cite{2018ApJ...862L..19N} \cite{2020MNRAS.498.5643T}; \cite{2021GCN.29375....1H}).

Soon after two neutron stars merge a relativistic outflow is produced by the merger remnant. This outflow travels outwards many orders of magnitude beyond the physical scale of the remnant into the circumstellar medium (CSM), while simultaneously expanding and accelerating as it converts internal energy into kinetic energy. After the outflow has swept up enough of the surrounding medium it begins to decelerate, which drives an external shock into the CSM. The magnetic field in the shocked region causes shock-accelerated electrons to spiral into helical, relativistic motion which produces broadband synchrotron radiation, known as the afterglow, which can remain detectable for months to years after the merger \citep[e.g.][]{2004RvMP...76.1143P}.

The afterglow light curve detected from GRB170817A has been used to learn about the GRB and its environment. Several different GRB models and parameter estimation techniques have been employed to constrain the radiation, hydrodynamic, and cosmological properties of the burst
(\cite{2018ApJ...869...55W}; \cite{2020ApJ...896..166R};  \cite{PhysRevLett.120.241103}; \cite{2019ApJ...886L..17H}; \cite{2019MNRAS.489.1919T}; \cite{2019ApJ...870L..15L}; \cite{2019NatAs...3..940H}; \cite{2019Sci...363..968G}; \cite{2022ApJ...938...12B}). In general it is agreed that GRB170817A was a relativistic structured jet with a characteristic opening angle, $\theta_{j}$, between 2$^{\circ}$ and 8$^{\circ}$ viewed off-axis at an angle, $\theta_{obs},$ between 15$^{\circ}$ and 35$^{\circ}$ (\cite{2021ApJ...909..114N}). An explanation for the difference in calculated jet geometries is given in \cite{2021ApJ...909..114N} where it was demonstrated that only the ratio of $\theta_{obs}/\theta_{j}$ can be constrained from the afterglow light curve alone and that additional information is needed to constrain each parameter separately. \cite{2018ApJ...865L...2Z} computed radio images from hydrodynamic simulations \citep{Xie2018} to demonstrate how the GRB170817A jet's viewing angle can be constrained with very long baseline interferometry (VLBI). \cite{2018Natur.561..355M} presented VLBI measurements of the radio source from 75 days to 230 days post-merger which alleviated model degeneracies and provided evidence for a collimated outflow with opening angle of 5$^{\circ}$ viewed at 20$^{\circ}$ off axis. Most recently, \cite{2022Natur.610..273M} use afterglow and VLBI measurements of superluminal motion to constrain the jet opening angle to $<5^{\circ}$ and the viewing angle to $19^{\circ}-25^{\circ}$.

In this paper we constrain properties of GRB170817A using broadband data and an eight-parameter afterglow model.  A two-parameter initial model is used to initiate hydrodynamic simulations of the relativistic outflow after as it expands into the CSM. Checkpoints containing full information about the fluid state are post-processed to calculate the synchrotron radiation, perform radiative transfer along different sight lines, and generate millions of different spectra. The synchrotron model described in \cite{1998ApJ...497L..17S} is used to fit the spectra and tabulate the spectral parameters needed to generate synthetic light curves from the hydrodynamics simulations. A light curve can be made in milliseconds and used in Markov-Chain Monte-Carlo analysis to find the best fitting light curves for a set of data and place constraints on the properties of the burst. The code used to generate light curves, fit observed data, and create parameter distributions has been compiled into a package called \texttt{JetFit} \footnote{https://github.com/NYU-CAL/JetFit} (\cite{Wu_2018}).

The viewing angle, $\theta_{obs}$, determined by this analysis is can be useful because of its ability to break the degeneracy between source inclination and absolute distance in the measured gravitational wave strain. The gravitational wave signal includes a combination of distance, sky position, inclination, chirp mass, and redshift. The product of chirp mass and redshift can be determined by the phase of the gravitational wave signal but additional information is needed to isolate the other properties (\cite{Nissanke_2010}). Electromagnetic detection of the source location and inclination information inferred by \texttt{JetFit} can be combined to reduce the uncertainty in the distance. An accurate determination of distance and redshift for a population of NS mergers could allow for a more precise calculation of cosmological parameters including the Hubble constant. 

In \S\ref{sec:jetfit} we describe the boosted fireball model in detail, the tabulation of spectral parameters, and the light curve generation. Previous results and the difference in reported values for $\theta_{obs}$ and $\theta_{j}$ are discussed in \S\ref{sec:prevwork}. In \S\ref{sec:results} we present and discuss the results of our model and we conclude in \S\ref{sec:conclusion} with a summary and discussion of future works.

\section{JetFit}\label{sec:jetfit}

\subsection{Boosted Fireball}
An outstanding challenge in modeling GRBs is picking a description of the energy of the jet as a function of angle. A simple description is the top-hat model where the jet energy, $\frac{dE}{d\Omega}$, is constant up to an opening angle, $\theta_c$, at which it drops to 0. However, this particular structure is not expected to occur naturally and the interpretation of afterglow properties, such as the observed jet-break, depends strongly on the choice of jet description (\cite{2002MNRAS.332..945R}). Semi-analytic models for the deceleration and spreading of the jetted outflow have been presented which attempt to explain the evolution of the jet in both 1D and 2D and compute afterglow radiation estimates from the hydrodynamics (\cite{2018Apj...865...94D}, \cite{2020arXiv200510313L}). Gaussian jets $\left(\frac{dE}{d\Omega} \propto e^{-\theta/\theta_c}\right)$ and power-law jets $\left(\frac{dE}{d\Omega} \propto (\theta/\theta_c)^{-p}\right)$ have also been proposed and used to fit data, however these models are guesses and have no physical motivation through first principles. 

A jet model with naturally occurring angular structure was introduced by \cite{2013ApJ...776L...9D} and is referred to as the "boosted fireball". The fireball represents a relativistic outflow that has been launched from the NS merger remnant and travelled many of orders of magnitude beyond the length scale of the remnant so that the details of its initial formation can be ignored. It is initialized with specific internal energy $\eta_0 \sim E/M$ and is given a bulk Lorentz factor of $\gamma_{B}$ with respect to the lab frame so that its total Lorentz factor, as measured in the lab frame, is $\Gamma \sim 2\gamma_B \eta_0$. 

After the fireball is launched, the dynamics of the outflow and resulting angular structure are completely determined by the two parameters $\eta_0$ and $\gamma_B$. In its center of momentum frame, the fireball expands isotropically as its internal energy is converted to kinetic energy until it reaches a maximum Lorentz factor of $\eta_0$ after which it becomes self-similar. After the fireball has swept up a mass of $\sim M/\eta_0$ it begins to decelerate which drives a shock wave into the medium; the fluid properties of this shock can be described by the Blandford-McKee solution (\cite{1976PhFl...19.1130B}). In the lab frame the fireball is beamed in the direction of the boost such that the outflow is confined to an angle $\theta_j \sim 1/\gamma_B$. This can be understood by considering the following: the fireball expands relativistically and the radius in its center of momentum frame is given by $R\sim\tau$ (in units of $c=1$), in the lab frame the expansion in the transverse direction is unaffected but along the boost the fireball has traveled a distance $d\sim \gamma_B\tau$, this creates an angle $\theta_j \sim R/d \sim 1/\gamma_B$. For low values of $\eta_0$ and $\gamma_B$ the fireball resembles a spherical outflow whereas high values create a jet-like outflow 



\begin{figure*}[t!]
        \includegraphics[width=\textwidth]{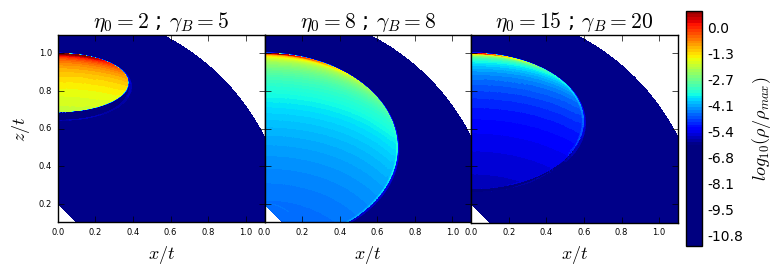}
        \caption{Examples of the structure of the boosted fireball for different values of $\eta_0$ and $\gamma_B$. The high density regions are shown in red. For higher values the material is focused into a thinner shell and a smaller angle about the z-axis. }
        \label{fig:BFcontour}
\end{figure*}


\cite{2013ApJ...776L...9D} showed that the angular structure of the boosted fireball could be written in terms of the isotropic energy 
\begin{equation}
E_{iso} \sim \frac{E}{1-cos(\theta_j)}
\end{equation}
and the maximum Lorentz factor
\begin{equation}
    \gamma_{max}(\theta) = \gamma_B \frac{\eta_0 + v_Bcos\theta\sqrt{\eta_0^2v_0^2 - \gamma_B^2v_B^2sin^2\theta}}{1+\gamma_B^2v_B^2sin^2\theta}
\end{equation}
where $v_0 = \sqrt{1-1/\eta_0^2}$ and $v_B$ and $v_{max}$ are the velocities associated with the boost and the local fluid velocity, respectively. The resulting angular structure is
\begin{equation}
\label{eq:BF_structure}
    \frac{dE}{d\Omega} = \frac{E_{iso}}{4\pi}\left(\frac{\gamma_{max}(\theta)}{\gamma_{max}(0)}\right)^3
\end{equation}

Figure \ref{fig:BFcontour} shows three examples of the density distribution for the boosted fireball model for different combinations of $\gamma_B$ and $\eta_0$. The numerical calculations are performed using the moving-mesh hydrodynamics code JET (\cite{2011ApJS..197...15D}, \cite{2013ApJ...775...87D}). In JET the computational zones move radially which allows for better capture of high Lorentz factor outflows such as the ones studied here.  

\subsection{Tabulating Spectra and MCMC}
 The radiative transfer is done by the \texttt{BoxFit} code detailed in \cite{2012ApJ...749...44V}. \texttt{BoxFit} uses the scale invariance of relativistic jets and their energy and circumburst density to greatly reduce the number of unique simulations needed to adequately explore the jet dynamics. Jets of almost arbitrary energy and density can be generated quickly once a basis of high-resolution simulations has been created. \texttt{BoxFit} uses the output from the boosted fireball hydrodynamic simulations and returns millions of spectra calculated from the outflows at different observer angles and intermediate $\eta_0$ and $\gamma_B$ values. The emission from the outflow is described using the standard synchrotron model in \cite{1998ApJ...497L..17S} which approximates each spectrum as a series of connected power laws. In this model each spectrum is completely parameterized by three spectral parameters $F_{peak}$, $\nu_m$, and $\nu_c$ which are the peak flux, minimum spectral frequency, and critical spectral frequency, respectively. We make use of the scaling relations in equation 4 of \cite{2015ApJ...799....3R} to write the observer time and three spectral parameters as functions of our model parameters. The full set of model parameters is $\{E_{iso}, n_0, \eta_0, \gamma_B; p, \epsilon_e, \epsilon_B, \xi_N; z, d_L, \theta_{obs}\}$ where $E_{iso}$ is the isotropic equivalent explosion energy, $n_0$ is the circumburst density, $p$ is the spectral index, $\epsilon_e$ is the electron energy fraction, $\epsilon_B$ is the magnetic energy fraction, $\xi_N$ is the fraction of accelerated electrons, $z$ is the redshift, and $d_L$ is the luminosity distance. In the following expressions, $f_{peak}, f_{m}, f_{c}$ are functions of the model parameters that can not be expressed analytically.:
 
\begin{equation}
t_{obs} = (1+z)\left(\frac{E_{iso}}{n_0}\right)^{1/3}\tau 
\end{equation}

\begin{equation}
    F_{peak} = \frac{1+z}{d^2_{L}} \frac{p-1}{3p-1}E_{iso} n^{1/2}_{0}\epsilon^{1/2}_B \xi_N f_{peak}(\tau; \eta_0, \gamma_B, \theta_{obs})
\end{equation}
     
 \begin{equation}    
     \nu_m = \frac{1}{1+z} \left(\frac{p-2}{p-1}\right)^2 n^{1/2}_0 \epsilon^2_e \epsilon^{1/2}_B \xi^{-2}_N f_m(\tau; \eta_0, \gamma_B, \theta_{obs}) \\
     \end{equation}

\begin{equation}     
     \nu_c = \frac{1}{1+z} E^{-2/3}_{iso} n^{-5/6}_0 \epsilon^{-3/2}_B \xi^{-2}_N f_c(\tau; \eta_0, \gamma_B, \theta_{obs})
 \end{equation}

 To approximate the functions $f_{peak}, f_{m},$ and $ f_{c}$, we fit each spectrum generated by \texttt{BoxFit} with the synchrotron model and calculate the values of the $f$ functions at particular values of $\tau, \eta_0, \gamma_B,$ and $\theta_{obs}$ and add them to a large table indexed by the four parameters. Once the table is completed for the full range of the indexing model parameters an arbitrary spectrum can be generated in milliseconds by specifying all the model parameters. This allows for rapid generation of light-curves and for MCMC fitting of the generated light-curves to observations. The fitting is done using the \texttt{emcee} package described in \cite{2013PASP..125..306F} which utilizes a group of parallel-tempered affine invariant walkers to explore the parameter space. 

\section{Previous Work \& Improvements}\label{sec:prevwork}

\cite{2018ApJ...869...55W} used MCMC with \texttt{JetFit} as the forward model along with with radio, optical, and X-ray afterglow data from \cite{2018ApJ...863L..18A} and \cite{2018ApJ...856L..18M} at $\sim$260 days post-merger to constrain the eight model parameters $\{E_{iso}, n_0, \eta_0, \gamma_B; p, \epsilon_e, \epsilon_B, \theta_{obs}\}$. In their analysis the hydrodynamics model parameter $\eta_0$ ranges from 2 to 10 and $\gamma_B$ ranges from 1 to 12; this range of parameter values corresponds to on-axis Lorentz factors of $\Gamma \sim 4 - 240$. Figure 3 in \cite{2018ApJ...869...55W} shows the one-dimensional projections of the posterior distributions for all eight model parameters. The posterior distributions for $\eta_0$ and $\gamma_B$ do not turn over or reach zero at the maximum of their range, indicating that there is non-zero probability at higher values. In this work we attempt to capture the posterior turn over by extending the parameters $\eta_0$ and $\gamma_B$ to 15 and 20, respectively. To cover the extended parameter space in $\eta_0$ and $\gamma_B$ we run additional boosted fireball simulations with \texttt{JET} (\cite{2013ApJ...775...87D}) and vary $\eta_0$ from 12 to 15 and $\gamma_B$ from 12 to 20 each in steps of 0.5.  This new hydrodynamic range allows us to fit afterglows from jets with Lorentz factors of up to $\Gamma \sim 600$.

\cite{2021ApJ...909..114N} review several publications that each use model fitting methods and afterglow data to determine the geometry of GW170817. The different studies agree that $\theta_{obs} \gg \theta_j$; despite this agreement the reported values for $\theta_{obs}$ and $\theta_j$ range from $14^{\circ} - 38^{\circ}$ and $2.5^{\circ} - 8^{\circ}$, respectively. Errors quoted in the works are several degrees for $\theta_{obs}$ and fractions of a degree for $\theta_j$. \cite{2021ApJ...909..114N} address the differences between measurements by showing that while the jet is relativistic there is a degeneracy in the shape of the light curve, $\theta_{obs}$, $\theta_j$, and $\Gamma$. This degeneracy means that a proper scaling of these three parameters can create afterglow light curves that look similar but are generated by different geometries. For this reason we report the posterior predictive distribution for $\theta_{obs}/\theta_j \sim\gamma_B\times \theta_{obs}$ when we discuss results in \S\ref{sec:results}. 


\section{Results and Discussion}
\label{sec:results}

\subsection{MCMC Analysis}

For our MCMC fitting we use radio, optical, and X-ray data from \cite{2022ApJ...927L..17H} up to $\sim$800 days after merger. We use similar transformations as \cite{Wu_2018} to enhance MCMC performance: $E_{j,50} \equiv E_j / 10^{50}$ ergs, $n_{0,0} \equiv n_0/1$ proton cm$^{-3}$ and $E_j / 10^{50}$, $n_{0,0}$, $\epsilon_e$, and $\epsilon_B$ are measured logarithmically. We can improve fitting results by setting the luminosity distance $d_L=39.5$ Mpc and redshift $z=0.0973$ in accordance with the values in the NASA Extragalactic Database for the source galaxy: NGC 4993. We denote the full parameter space as $\Theta$:
\begin{equation}
    \Theta = \{\log_{10}E_{0,50}, \log_{10}n_{0,0}, \eta_0, \gamma_B, \theta_{obs}, \log_{10}\epsilon_e, \log_{10}\epsilon_B, p\}
\end{equation}
For all priors except $\theta_{obs}$ we use a uniform distribution with bounds chosen to cover phenomenologically relevant values. The prior used for $\theta_{obs}$ is sin$(\theta_{obs})$ to account for geometric observational effects. The prior bounds used are given in Table (\ref{tab:prior_bounds}).

The Python package \texttt{emcee} is used to sample the posteriors of the parameters listed in $\Theta$. We expect the parameter space to contain multi-modal distributions and therefor utilize the parallel-tempered sampling method available in the package. For each fitting we use 100 walkers and 10 different temperatures. Each run has a burn-in period of 6000 steps and is then ran for an additional 50000 steps. For each of the following combinations of parameters and data we run \texttt{emcee} several times to confirm convergence. Additionally, we compute an estimate for the autocorrelation time for each parameter to guarantee that enough independent samples have been taken.

\begin{table}
    \centering
    \begin{tabular}{c c}
        Parameter & Range \\
        \hline \hline
        $\log_{10}E_{j, 50}$ & [-6, 3] \\
        $\log_{10}n_{0,0}$ & [-6, -3] \\
        $\eta_0$ & [2, 15] \\ 
        $\gamma_B$ & [1, 20] \\
        $\theta_{obs}$ & [0, 1] \\
        $\log_{10}\epsilon_B$ & [-6, 0] \\ 
        $\log_{10}\epsilon_e$ & [-6, 0] \\ 
        $p$ & [2, 2.5] \\
    \end{tabular}
    \caption{Bounds for prior distributions}
    \label{tab:prior_bounds}
\end{table}

In order to compare our updated model to the previous work, we initially fit for all parameters included in $\Theta$ and use only the broadband data used in \cite{Wu_2018}. A comparison between the parameter values found in the two works is presented in Table (\ref{tab:comparison_all}). Next, we fix the circumburst density and perform the fit again. Comparisons between values found by \cite{Wu_2018} and this work are in Table (\ref{tab:comparison_nconst}). From these comparisons we see that the parameter values found by each model are all within the stated uncertainty of each other. We proceed by fitting the most recent afterglow data with our expanded model. 

\begin{table}
    \centering
    \begin{tabular}{c c c}
        Parameter & Median [YW] & Median [AM] \\
        \hline \hline
        $\log_{10}E_{j, 50}$ & $0.04^{+1.36}_{-0.98}$ & $0.33^{+1.21}_{-0.99}$ \\ 
        $\log_{10}n_{0, 0}$ & $-1.4^{+1.4}_{-1.2}$ & $-1.30^{+1.28}_{-1.51}$ \\ 
        $\eta_0$ & 8.0$^{+1.88}_{-0.94}$ & $8.30^{+4.03}_{-2.09}$ \\
        $\gamma_B$ & 11.06 & $9.58^{+5.17}_{-3.26}$ \\
        $\theta_{obs}$ & 0.47$^{+0.17}_{-0.05}$ & $0.55^{+0.24}_{-0.22}$\\
        $\log_{10}\epsilon_e$ & -0.65$^{+0.49}_{-1.87}$ & $-1.53^{+0.96}_{-1.40}$ \\ 
        $\log_{10}\epsilon_B$ & -5.9$^{+2.4}_{-0.0}$ & $-4.47^{+1.60}_{-1.07}$ \\ 
        $p$ & 2.154$^{+0.012}_{-0.010}$ & $2.16^{+0.01}_{-0.01}$\\ 
        $\gamma_B \times \theta_{obs}$ & 5.19$^{+3.12}_{-1.69}$ & 5.06$^{+0.53}_{-0.44}$ \\
    \end{tabular}
    \caption{Comparison between previous analysis done by \cite{Wu_2018} [YW] and current results [AM]. Medians and their bounds are found with MCMC sampling using afterglow data reported in \cite{2018ApJ...863L..18A} and \cite{2018ApJ...856L..18M}. Uncertainties in $\gamma_B$ are not reported in \cite{Wu_2018}.} 
    \label{tab:comparison_all}
\end{table}

\begin{table}
    \centering
    \begin{tabular}{c c c}
        Parameter & Median [YW] & Median [AM] \\
        \hline \hline
        $\log_{10}E_{j, 50}$ & $-0.81^{+0.26}_{-0.39}$ & $-0.50^{+1.18}_{-0.65}$ \\ 
        $\eta_0$ & 8.0 & $7.71^{+4.49}_{-1.92}$ \\
        $\gamma_B$ & 11.11 & $12.50^{+5.13}_{-2.05}$ \\
        $\theta_{obs}$ & 0.47$^{+0.08}_{-0.03}$ & $0.41^{+0.10}_{-0.15}$\\
        $\log_{10}\epsilon_e$ & -0.51$^{+0.35}_{-0.75}$ & $-0.92^{+0.52}_{-1.01}$ \\ 
        $\log_{10}\epsilon_B$ & -1.91$^{+0.3}_{-1.18}$ & $-2.95^{+0.99}_{-1.62}$ \\ 
        $p$ & 2.154$^{+0.012}_{-0.012}$ & $2.16^{+0.01}_{-0.01}$\\ 
        $\gamma_B \times \theta_{obs}$ & 5.22$^{+2.15}_{-1.31}$ & 5.13$^{+0.58}_{-0.87}$ \\
    \end{tabular}
    \caption{Comparison between previous analysis done by \cite{Wu_2018} [YW] and current results [AM] with $\log_{10}n_{0,0}=10^{-3}$. Medians and their bounds are found with MCMC sampling using afterglow data reported in \cite{2018ApJ...863L..18A} and \cite{2018ApJ...856L..18M}. Uncertainties in $\gamma_B$ and $\eta_0$ are not reported in \cite{Wu_2018}.}
    \label{tab:comparison_nconst}
\end{table}

\begin{figure}
        \includegraphics[width=\columnwidth]{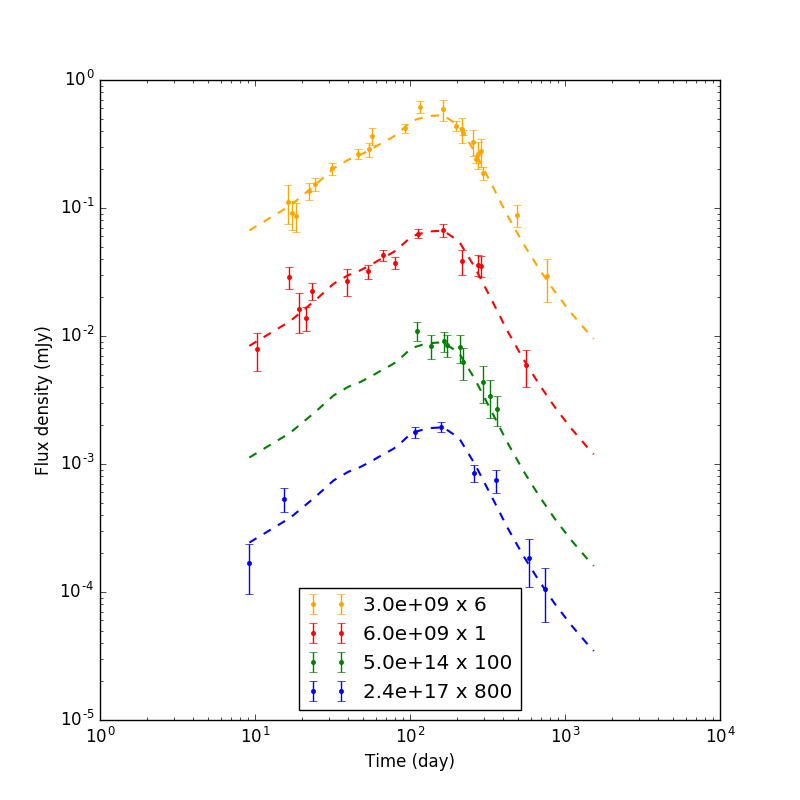}
        \caption{Best fitting light curves from MCMC results for full $\Theta$ parameter space using data up to $\sim800$ days post merger. The inset labels observational frequency and a scale factor used for visualization purposes.}
        \label{fig:lc_all}
\end{figure}

\begin{figure*}[t!]
    \includegraphics[width=\textwidth]{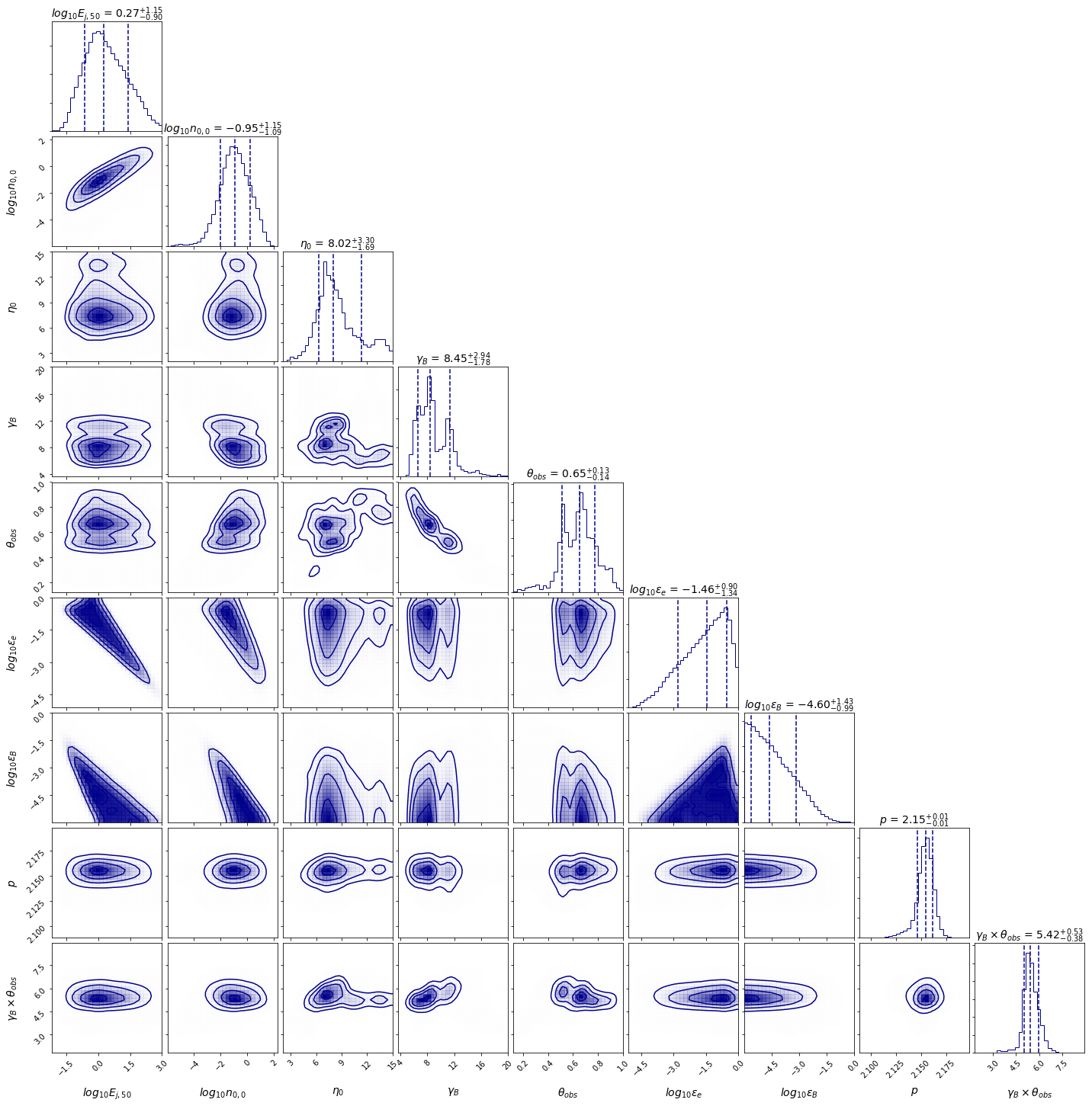}
       \caption{Contours and 1D distributions from posterior projections of MCMC results with full $\Theta$ parameter space using data up to $\sim800$ days post merger.}
\label{fig:contour_all}
\end{figure*}

We find that the rise, peak, and fall of the light curves are all captured adequately well as shown in Figure (\ref{fig:lc_all}). The 2D contour plots and their 1D marginalization are shown in Figure (\ref{fig:contour_all}) and the median values for each parameter are listed in Table (\ref{tab:median}). The data prefer a relativistic outflow viewed off-axis with $E_{j}\sim10^{50}$ ergs, $\eta_0\sim8$, and $\gamma_B\times\theta_{obs}\sim5.4$. The median value we find for $\gamma_B\times\theta_{obs}$ is in agreement with values found by other studies as summarized by \cite{2021ApJ...909..114N}. Additionally, the lower uncertainty in the circumburst density, $\log_{10}n_0$, is consistent with the observational constraints placed in \cite{2019ApJ...886L..17H}. For some pairs of parameters, a re-scaling will produce similar forward models as long as the ratio of the two parameters is unchanged. Degeneracies between model parameters are evident in the 2D contour plots. Specifically, the circumburst medium density $n$ is degenerate with $E_j$, $\epsilon_B$, and $\epsilon_e$ and the observer angle $\theta_{obs}$ is degenerate with $\gamma_B$. 

\begin{table}
    \centering
    \begin{tabular}{c  c}
        Parameter & Median \\
        \hline \hline
        $\log_{10}E_{j, 50}$ & $0.27^{+1.15}_{-0.9}$ \\ 
        $\log_{10}n_{0, 0}$ & $-0.95^{+1.15}_{-1.09}$ \\ 
        $\eta_0$ & 8.02$^{+3.30}_{-1.69}$ \\
        $\gamma_B$ & 8.45$^{+2.94}_{-1.78}$ \\
        $\theta_{obs}$ & 0.65$^{+0.13}_{-0.14}$ \\
        $\log_{10}\epsilon_e$ & -1.46$^{+0.90}_{-1.34}$ \\ 
        $\log_{10}\epsilon_B$ & -4.60$^{+1.43}_{-0.99}$ \\ 
        $p$ & 2.15$^{+0.01}_{-0.01}$ \\ 
        $\gamma_B \times \theta_{obs}$ & 5.42$^{+0.53}_{-0.38}$ \\
    \end{tabular}
    \caption{Median parameter values for full $\Theta$ using data up to $\sim800$ days post merger.}
    \label{tab:median}
\end{table}

\begin{figure*}[t!]
    \includegraphics[width=\textwidth]{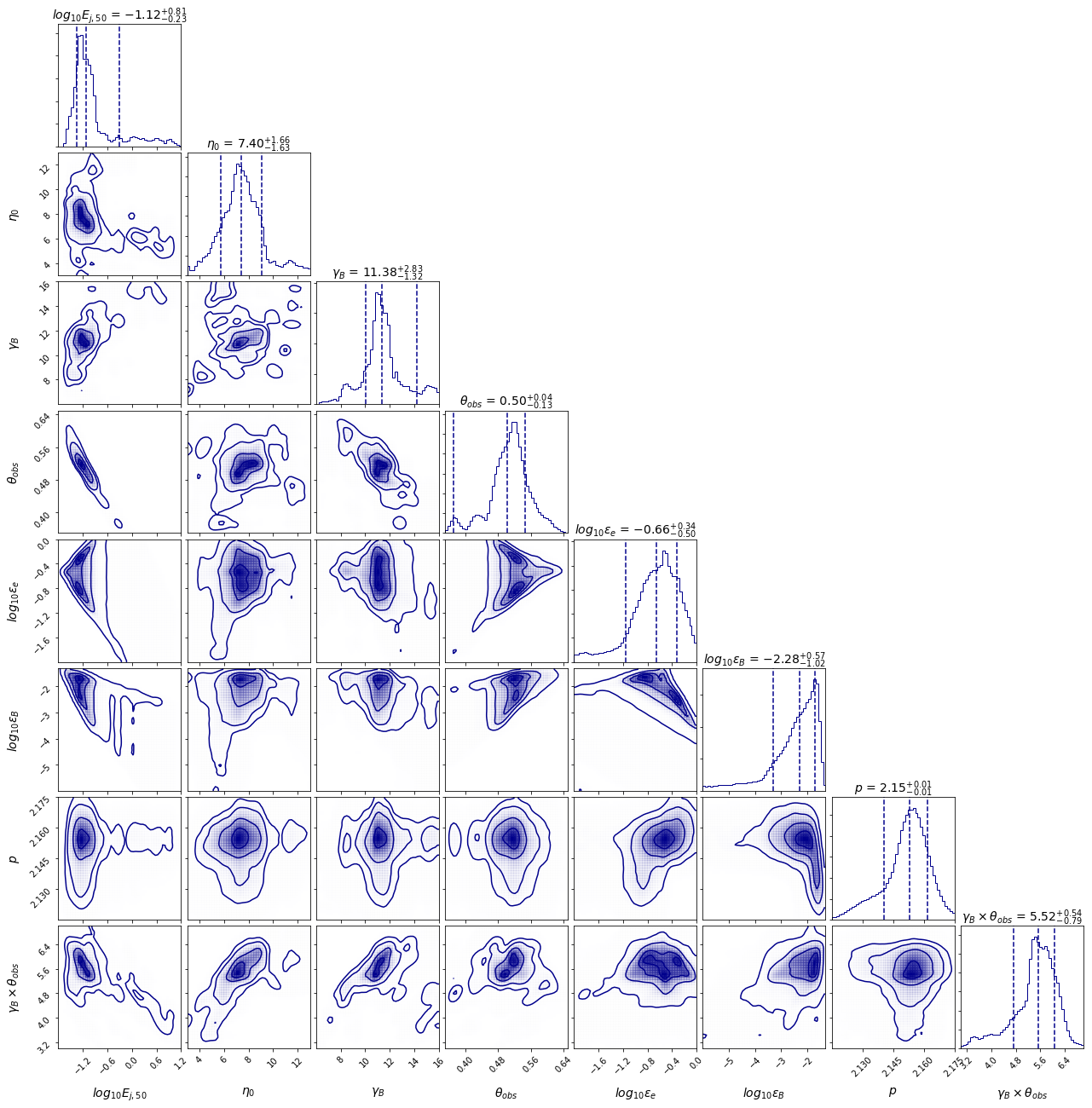}
       \caption{Contours and 1D distributions from posterior projections of MCMC results with a constant $n_{0,0}$ using data up to $\sim800$ days post merger.}
\label{fig:contour_nconst}
\end{figure*}

To further reduce dimensionality and improve constraints on the parameters we fix $n_{0,0}=9.6\times10^{-3}$ to match observational constraints set in \cite{2019ApJ...886L..17H}. The best fitting light curves when the density is constant are identical to light curves shown in Figure (\ref{fig:lc_all}). Corner plots featuring 2D and 1D projections of posterior distributions are shown in Figure (\ref{fig:contour_nconst}) and their median values are listed in Table (\ref{tab:median_nconst}). We find that all 1D posterior distributions peak within the bounds listed in Table (\ref{tab:prior_bounds}) and there are no indications of distributions with significant probability at higher, unexplored values. The data prefer a relativistic fireball with $E_j\sim7.5 \times 10^{48}$ ergs, $\eta_0\sim7.4$, $\gamma_B\sim11.4$, and $\theta_{obs}\sim0.5$ radians. Furthermore, we find agreement in both our jet opening angle, $\theta_j \sim 1/\gamma_B$, and viewing angle with the results presented in \cite{2022Natur.610..273M} ($\theta_j=4.87^{+1.83}_{-1.14}$ deg, $\theta_{obs}=21.98^{+3.35}_{-2.89}$ deg) which include superluminal motion measured by very long baseline interferometry. 

\begin{table}
    \centering
    \begin{tabular}{c c}
        Parameter & Median \\
        \hline \hline
        $\log_{10}E_{j, 50}$ & $-1.12^{+0.81}_{-0.23}$ \\ 
        $\eta_0$ & 7.40$^{+1.66}_{-1.63}$ \\
        $\gamma_B$ & 11.38$^{+2.83}_{-1.32}$ \\
        $\theta_{obs}$ & 0.50$^{+0.04}_{-0.13}$ \\
        $\log_{10}\epsilon_B$ & -0.66$^{+0.34}_{-0.50}$ \\ 
        $\log_{10}\epsilon_e$ & -2.28$^{+0.57}_{-1.02}$ \\ 
        $p$ & 2.15$^{+0.01}_{-0.01}$ \\
        $\gamma_B \times \theta_{obs}$ & 5.52$^{+0.54}_{-0.79}$ \\
    \end{tabular}
    \caption{Median parameter values for constant density}
    \label{tab:median_nconst}
\end{table}

\begin{figure*}[t!]
    \includegraphics[width=\textwidth]{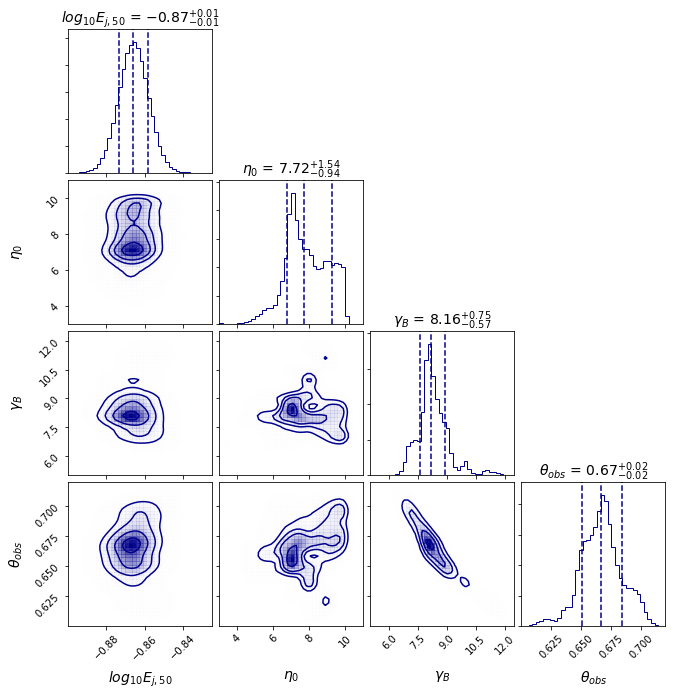}
       \caption{1D and 2D posterior projections for geometric parameters with fixed $n_{0,0}$, $\epsilon_B$, $\epsilon_e$, $p$}
\label{fig:contour_geom}
\end{figure*}

Lastly we fit only for the geometry of the jet by fixing the values of $\epsilon_B$, $\epsilon_e$, and $p$ to their median values listed in Table (\ref{tab:median}). The resulting posteriors and their projections are shown in Figure (\ref{fig:contour_geom}). In this case the data prefer a slightly more relativistic jet with $E_j\sim10^{49}$, $\eta_0\sim7.7$, $\gamma_B\sim8$, and an observer angle of $\theta_{obs}\sim0.67$. We see a noticeable degeneracy between $\gamma_B$ and $\theta_{obs}$ which is in accordance with the argument presented in \cite{2021ApJ...909..114N} about the inherent degeneracy in the shape of afterglow light curves. 

Each analysis mentioned above shows that the data prefer a relativistic, structured jet that is viewed off-axis. The range of parameter values found in this work overlap with those reported in \cite{Wu_2018} even with the addition of new late-time afterglow data and a larger parameter space. These findings reinforce the validity of the previous results and show that a narrower, on-axis jet is not preferred by the data even when the parameter space is extended to include such models. 

\subsection{Implications for Cosmology}
\label{sec:cosmo_imps}

Gravitational waves from compact object mergers can be used to measure the expansion rate of the universe, $H(z)$ (\cite{1986Natur.323..310S}). The gravitational waves act as a standard siren, similar to the concept of standard candles, and can be used to infer the luminosity distance to the source independent of any distance ladder (\cite{2022PhRvL.129f1102E}). Gravitational wave sources alone are generally poorly localized, however, an electromagnetic counterpart can reduce localization error and may provide the source's redshift. The merger of two neutron stars is therefore of particular utility because its production of gravitational waves, a short-GRB, and afterglow radiation can be used to measure the Hubble constant.  

It has been shown in \cite{2010ApJ...725..496N} that the measured gravitational wave strain can be expressed as a linear combination of two polarizations, $h_+$ and $h_{\times}$. Each polarization contains terms that depend on the cosine of the angle of inclination of the neutron star binary's orbital plane to the observer's line of sight. The \texttt{JetFit} analysis puts constraints on this viewing angle, $\theta_{obs}$, which helps to alleviate the degeneracy between the source's absolute distance and the inclination as measured by the gravitational wave strain.  \cite{2021ApJ...909..218A} report a measurement of $H_0$ that was found using the distance inferred from the gravitational wave signal from GW170817 and the local Hubble flow velocity at the position of the host galaxy, NGC 4993. Their Bayesian analysis produces a posterior distribution on $H_0$ and the source inclination which could be improved by including our constraints on $\theta_{obs}$ into their priors. As more multi-messenger events are detected, robust estimates for the inclination angle will be helpful in constraining the Hubble constant. The \texttt{JetFit} tool provides a thorough exploration of the possible geometries and can therefore improve constraints on $H_0$ estimates. 

\section{Conclusions}
\label{sec:conclusion}
In this work we used the most recent broadband afterglow observations of GW170817 and parallel-tempered, affine invariant MCMC fitting of a parameterized model to data to constrain hydrodynamic, radiative, and cosmological properties of the associated BNS merger: GW170817. We find that when fitting with the full model $\Theta$ the data prefer a relativistic outflow viewed off-axis with $E_j\sim10^{50}$ ergs, $\eta_0\sim8$ and $\gamma_B\times\theta_{obs}\sim5.4$. 

As it evolves, the afterglow flux will become dominated by longer wavelengths and future observations will take place mainly in the radio band. Synchrotron self-absorption becomes important at frequencies $\sim  6$ GHz and will need to be incorporated into the spectral fitting process in future works. 

Although GW170817 is currently a unique event, it is expected that the LIGO-VIRGO network will detect several more NS-NS mergers as they approach maximum sensitivity. The mergers can act as standard sirens and open the possibility of making measurements of $H_0$ as more are detected.

\acknowledgements
We acknowledge support from NASA grant 21-ATP21-0108. We thank David Hogg for his feedback on our \texttt{emcee} results.

\nocite{*}
\bibliography{ms}{}
\bibliographystyle{aasjournal}
\end{document}